\documentclass[conference]{IEEEtran}
\IEEEoverridecommandlockouts
\usepackage{cite}
\usepackage{amsmath,amssymb,amsfonts}
\usepackage{algorithmic}
\usepackage{graphicx}
\usepackage{textcomp}
\usepackage{xcolor}
\def\BibTeX{{\rm B\kern-.05em{\sc i\kern-.025em b}\kern-.08em
    T\kern-.1667em\lower.7ex\hbox{E}\kern-.125emX}}
\begin{document}

\title{Unshuffling fields in data formats\\
\thanks{This material is based upon work partially supported by the Defense Advanced Research Projects Agency (DARPA) SafeDocs program under contract HR001119C0072. Any opinions, findings and conclusions or recommendations expressed in this material are those of the author(s) and do not necessarily reflect the views of DARPA.}
}

\author{\IEEEauthorblockN{Steve Huntsman}
\IEEEauthorblockA{\textit{Cyber Technology} \\
\textit{BAE Systems FAST Labs}\\
4301 North Fairfax Drive, Suite 800\\
Arlington, Virginia, USA \\
steve.huntsman@baesystems.com}
}

\maketitle

\begin{abstract}
Data format reverse engineering commonly involves identifying conserved format motifs. However, this process typically requires establishing a common ordering for format elements across instances, particularly for formats using type-(length)-value tuples or ``chunk'' encoding. It is useful to \emph{unshuffle} chunks with common length statistics as a precursor to identifying conserved internal structures. We formalize the unshuffling problem and subsequently derive probabilistic bounds and outline corresponding algorithms for it. We empirically demonstrate unshuffling and highlight connections with the related class of synchronization problems.
\end{abstract}

\begin{IEEEkeywords}
format reverse engineering, information theory, type-length-value encoding, synchronization
\end{IEEEkeywords}

\section{\label{sec:Introduction}Introduction}

Many data formats are structured according to \emph{type-length-value} (TLV) encoding \cite{McCannChandra,MedhiRamasamy}, wherein the semantic type of some data is specified by a short field (typically a few bits or bytes), the length of the (possibly padded) data value is encoded in a fixed-length field, and the data value itself is in a variable-length field. TLV encoding---or, roughly equivalently, so-called ``chunk'' formatting---allows data to be parsed without explicit delimiters or even a fixed ordering. 

For example, the Data Distribution Service (DDS) platform \cite{DDS} for Internet of Things (IoT) applications (e.g., transportation, energy, medical devices, smart cities, military/aerospace, industrial control [SCADA] systems, etc.) uses the RTPS wire protocol \cite{RTPS}. RTPS in turn carries messages comprised of submessages that use TLV encoding and that carry still additional internal structure of predefined submessage ``elements.'' Meanwhile, an actual application layer encapsulated in DDS/RTPS can be essentially arbitrary: the encapsulated data/payloads are serialized and uninterpreted by RTPS. As IoT devices continue to proliferate, idiosyncratic and/or undocumented DDS/RTPS-encapsulated applications and their ilk will too, and reverse engineering these applications' protocols will become increasingly important to ensure the safety and security of integrated IoT technologies.

Towards that end, this paper concerns itself with the problem of identifying and \emph{unshuffling} chunks in data formats using very weak assumptions. This practice (demonstrated in later figures) is useful for format/protocol reverse engineering \cite{Duchene,Kleber}: for example, one might try to use multiple alignment techniques to identify conserved structures in a data corpus (see, e.g., \cite{Netzob}), but this tactic fails unless chunks occur in comparable order.

We assume that chunks may occur in any order, but that the chunks in format exemplars have identical conserved internal structure (e.g., fixed bytes at certain positions) and lengths (including multiplicity). This latter assumption is justified by focusing on exemplars within a large corpus that have a fixed short length and clustering the results according to an appropriate dissimilarity measure (e.g., Hellinger distance of byte distributions or relative de Bruijn entropy of byte sequences \cite{Huntsman}). Such a tactic allows us to isolate files or messages that are likely in practice to have chunks whose syntax (versus embedded data) is equivalent, and so we restrict consideration to this situation in the paper. The general case can be addressed by recognizing the resulting observed syntax.

In general, we write $M$ for the number of chunks in a file or message, and $(L_1,\dots,L_M)$ to indicate the fixed lengths of chunks within a corpus of files or messages (obtained by isolation along the lines above if necessary). Informally, we are concerned with the problem where we have $N$ files or messages, each containing $M$ chunks in some arbitrary order. We derive probabilistic bounds and outline corresponding algorithms for reconstructing block permutations of the chunks that bring them into a common order. 

The paper is organized as follows. \S \ref{sec:2Unshuffling} discusses the case $M = 2$ in detail; \S \ref{sec:MUnshuffling} addresses the case of generic $M$ and side conditions that make the problem tractable, and \S \ref{sec:SynchronizationProblems} discusses the interpretation of the unshuffling problem as a special instance of a so-called \emph{synchronization problem}. Finally, \S \ref{sec:Permutations} details the basic algebra of block permutations necessary to formalize the general problem and arguments.

\section{\label{sec:2Unshuffling}2-unshuffling}

\subsection{\label{sec:2UnshufflingProblem}Problem}

Suppose that we have a ``template'' tuple with components in a finite set and which is repeatedly sampled after first randomizing a fixed subset of components and occasionally cyclically shifting the resulting components by a fixed offset. The general 2-unshuffling problem is roughly to determine which samples have been offset, to reconstruct the value of the offset, and subsequently to align the samples with the underlying template via (appropriate block) cyclic shifts.

A formal description of the problem setup follows. For convenience, we shall write $\bar y := 1-y$. Let $1 < q \in \mathbb{N}$; let $1 \ll L_m \in \mathbb{N}$ for $m \in \{1,2\}$ and write $L := L_1 + L_2$; let $\pi := (L_1+1,\dots,L,1,\dots,L_1) \in S_L$, where the symmetric group on $L$ elements is indicated. Let $\mathcal{L} \subset [L] := \{1,\dots, L\}$ be uniformly random for $\lambda := |\mathcal{L}|/L$ fixed and define the random variable $\xi^{(\mathcal{L})} \in (\mathbb{Z}/q\mathbb{Z})^L$ by 
\footnote{
NB. For the most part, we could replace $\mathbb{Z}/q\mathbb{Z}$ with $[q]$ in our discussion: however, we retain the former so as to cleanly consider ``additive noise'' as in \eqref{eq:A}.
}
\begin{equation}
\label{eq:xi}
\xi^{(\mathcal{L})}_\ell \sim
\begin{cases}
\mathcal{\mathcal{U}}(\mathbb{Z}/q\mathbb{Z}), & \ell \in \mathcal{L} \\
0, & \ell \notin \mathcal{L}.
\end{cases}
\end{equation}
Here we write $\xi \sim \mathcal{U}(\Omega)$ to indicate that $\xi$ is uniformly distributed on $\Omega$. We shall write $\xi^{(\mathcal{L},n)}$ for the $n$th sample of $\xi^{(\mathcal{L})}$ (note that the samples $\xi^{(\mathcal{L},n)}$ are necessarily IID). Finally, for $1 \ll N \in \mathbb{N}$, let $\mathcal{N} \subset [N]$ be uniformly random for $\nu := |\mathcal{N}|/N$ fixed (we may take $\nu \le 1/2$ without loss of generality), let $x_\ell \sim \mathcal{U}(\mathbb{Z}/q\mathbb{Z})$ for $\ell \in [L]$, and define the random matrix $A = A^{(x,\mathcal{L},\mathcal{N})} \in M_{L,N}(\mathbb{Z}/q\mathbb{Z})$ by
\begin{equation}
\label{eq:A}
A_{\ell n} = A^{(x,\mathcal{L},\mathcal{N})}_{\ell n} :=
\begin{cases}
\left ( x+\xi^{(\mathcal{L},n)} \right )_{\pi(\ell)}, & n \in \mathcal{N} \\
\left ( x+\xi^{(\mathcal{L},n)} \right )_\ell, & n \notin \mathcal{N}.
\end{cases}
\end{equation}
Writing $\alpha_\ell(n) := \left ( x + \xi^{(\mathcal{L},n)} \right )_\ell$, we have $A_{\ell n} = 1_\mathcal{N}(n) \alpha_{\pi(\ell)}(n) + (1-1_\mathcal{N}(n)) \alpha_\ell(n)$.

The problem is now this: given $A$, find $\mathcal{N}$ and $\pi$ (and to the extent that it matters, $\mathcal{L}$).

\subsection{\label{sec:2UnshufflingN}Finding $\mathcal{N}$}

\textsc{Lemma.}
For $0 < \nu < 1$, the probability $p_\mathcal{N}$ that the function $n \mapsto A_{\ell n}$ takes exactly two values with inverse images $\mathcal{N}$ and $[N] \backslash \mathcal{N}$ is given by
\begin{equation}
\frac{p_\mathcal{N}}{1-q^{-1}} = \bar \lambda^2 + \bar \lambda \lambda (q^{1-N\nu}+q^{1-N\bar \nu}) + \lambda^2 q^{2-N}.
\end{equation}
In particular, for $0 < \nu < 1$ and $N \rightarrow \infty$, 
\begin{equation}
p_\mathcal{N} \sim (1-q^{-1}) \cdot \bar \lambda^2.
\end{equation}

\textsc{Proof.}
Write $X_0 := \alpha_\ell([N] \backslash \mathcal{N})$ and $X_1 := \alpha_{\pi(\ell)}(\mathcal{N})$. Write $E_{\ne}$ for the event $X_0 \ne X_1$ and write $E_m$ for the event $|X_m| = 1$ for $m \in \{0,1\}$. The probability we want to compute is $\mathbb{P}(E_{\ne} \cap E_0 \cap E_1)$. It is obvious that $\mathbb{P}(E_{\ne} | E_0 \cap E_1) = 1-q^{-1}$, and furthermore that $E_0$ and $E_1$ are independent. As such, in order to compute $\mathbb{P}(E_0 \cap E_1)$, first consider the probability that the function $\alpha_\ell$ is constant on a set of $r$ elements. If $\ell \in \mathcal{L}$, this probability is $q^{1-r}$ and any of the elements of $\mathbb{Z}/q\mathbb{Z}$ are possible values of $\alpha_\ell$ on the set in question, whereas if $\ell \notin \mathcal{L}$, this probability is $1$ and $\alpha_\ell = x_\ell$ on the set in question. With this in mind, consider the following table:
\begin{center}
\begin{tabular}{ c | c | c | c | c }
	$\ell \in \mathcal{L}$ & $\pi(\ell) \in \mathcal{L}$ & weight & $\mathbb{P}(E_0)$ & $\mathbb{P}(E_1)$ \\ \hline 
	\rule{0pt}{3ex}    $\bot$ & $\bot$ & $\bar \lambda \bar \lambda$ & $1$ & $1$ \\ 
	$\bot$ & $\top$ & $\bar \lambda \lambda$ & $1$ & $q^{1-N \nu}$ \\
	$\top$ & $\bot$ & $\lambda \bar \lambda$ & $q^{1-N \bar \nu}$ & $1$ \\
	$\top$ & $\top$ & $\lambda \lambda$ & $q^{1-N \bar \nu}$ & $q^{1-N \nu}$ \\
\end{tabular}
\end{center}
The lemma now immediately follows from the information in this table and the preceding observations. $\Box$

\

A similar but slightly more delicate calculation yields the following

\

\textsc{Lemma.}
The probability $p_2$ that the function $n \mapsto A_{\ell n}$ takes exactly two values is given by 
\begin{align}
\frac{p_2-p_\mathcal{N}}{2(1-q^{-1})} & = & ( \bar \lambda \lambda + \lambda^2 \cdot q^{1-N\bar \nu} ) \cdot (2^{N \nu-1}-1) q^{1-N \nu} \nonumber \\
& + & ( \lambda \bar \lambda + \lambda^2 \cdot q^{1-N \nu} ) \cdot (2^{N \bar \nu-1}-1) q^{1-N \bar \nu} \nonumber \\
& + & \lambda^2 \cdot (2^{N \bar \nu-1}-1) (2^{N \nu-1}-1) q^{2-N}. 
\end{align}

\textsc{Proof.}
Extending the notation introduced in the previous lemma, write $E_\subset$ for the event $X_0 \subset X_1$, etc. and $E'_m$ for the event $|X_m| = 2$ for $m \in \{0,1\}$. The desired probability is the sum of the probabilities of the four disjoint events $E_{\neq} \cap E_0 \cap E_1$, $E_\subset \cap E_0 \cap E'_1$, $E_\supset \cap E'_0 \cap E_1$, and $E_= \cap E'_0 \cap E'_1$. The previous lemma gives an expression for the first of these events. Moreover, $\mathbb{P}(E_\subset | E_0 \cap E'_1) = \mathbb{P}(E_\supset | E'_0 \cap E_1) = 2/q$ and $\mathbb{P}(E_= | E'_0 \cap E'_1) = 1/\binom{q}{2}$.

Consider now the probability that the function $\alpha_\ell$ takes $s$ values over a set of $r$ elements. If $\ell \in \mathcal{L}$, this probability is $S(r,s) \cdot \binom{q}{s} s! \cdot q^{-r}$, where $S(\cdot,\cdot)$ denotes a Stirling number of the second kind \cite{Stanley}: for $s = 2$, this probability is  $(2^{r-1}-1) (q-1) q^{1-r}$. If $\ell \notin \mathcal{L}$, this probability is $\delta_{v1}$. With this in mind, consider the following table:
\begin{center}
\begin{tabular}{ c | c | c | c | c }
	$\ell \in \mathcal{L}$ & $\pi(\ell) \in \mathcal{L}$ & weight & $\mathbb{P}(E'_0)$ & $\mathbb{P}(E'_1)$ \\ \hline 
	\rule{0pt}{3ex}    $\bot$ & $\bot$ & $\bar \lambda \bar \lambda$ & $0$ & $0$ \\ 
	$\bot$ & $\top$ & $\bar \lambda \lambda$ & $0$ & $p'_1$ \\
	$\top$ & $\bot$ & $\lambda \bar \lambda$ & $p'_0$ & $0$ \\
	$\top$ & $\top$ & $\lambda \lambda$ & $p'_0$ & $p'_1$ \\
\end{tabular}
\end{center}
Here, $p'_0 := (2^{N \bar \nu-1}-1) (q-1) q^{1-N \bar \nu}$ and $p'_1 := (2^{N \nu-1}-1) (q-1) q^{1-N \nu}$.

It follows that $\mathbb{P}(E_\subset \cap E_1 \cap E'_2)$, $\mathbb{P}(E_\supset \cap E'_1 \cap E_2)$, and $\mathbb{P}(E_= \cap E'_1 \cap E'_2)$ respectively equal
\begin{equation}
\frac{2}{q} \cdot ( \bar \lambda \lambda + \lambda^2 \cdot q^{1-N\bar \nu} ) \cdot (2^{N \nu-1}-1) (q-1) q^{1-N \nu}, \nonumber
\end{equation}
\begin{equation}
\frac{2}{q} \cdot ( \lambda \bar \lambda + \lambda^2 \cdot q^{1-N \nu} ) \cdot (2^{N \bar \nu-1}-1) (q-1) q^{1-N \bar \nu}, \nonumber
\end{equation}
and
\begin{equation}
\frac{2}{q(q-1)} \cdot \lambda^2 \cdot (2^{N \bar \nu-1}-1) (2^{N \nu-1}-1) (q-1)^2 q^{2-N}. \nonumber
\end{equation}
The lemma now immediately follows from the information in this table and the preceding observations. $\Box$

\

We therefore have that for $0 < \nu < 1$ and $N \rightarrow \infty$, $p_2/(1-q^{-1})$ asymptotically approaches
\begin{equation}
\bar \lambda^2 + 2\bar \lambda \lambda \left [ \left ( \frac{q}{2} \right )^{1-N \nu} + \left ( \frac{q}{2} \right )^{1-N \bar \nu} \right ] + 2 \lambda^2 \left ( \frac{q}{2} \right )^{2-N} \nonumber
\end{equation}
and we obtain the following

\

\textsc{Theorem.}
For $0 < \nu < 1$, the probability that the function $n \mapsto A_{\ell n}$ takes two values with inverse images \emph{not} $\mathcal{N}$ and $[N] \backslash \mathcal{N}$ is $p_2 - p_\mathcal{N} = O \left ( (q/2)^{-N \min(\bar \nu, \nu)} \right )$. $\Box$

\

Thus with high probability, we can identify $\mathcal{N}$ by simply considering the partitions of $[N]$ induced by the inverse images of the functions corresponding to rows of $A$: if $q > 2$, the vast majority of such partitions with two elements will be $\{\mathcal{N},[N] \backslash \mathcal{N}\}$. If $q = 2$, we can still identify $\mathcal{N}$ with high probability by focusing on the most frequently occurring partition (and of course we ought to do this anyway in applications), though we will not attempt to produce quantitative estimates for this case.

\subsection{\label{sec:2UnshufflingPi}Finding $\pi$}

Given an estimate $\hat{\mathcal{N}}$ of $\mathcal{N}$ that is exact with high probability, the question now becomes how to estimate $\pi$ (equivalently, the RHS of $L = L_1 + L_2$). Define 
\begin{eqnarray}
\hat{\mathcal{L}}_0 & := & \{ \ell \in [L] : | \{A_{\ell n} : n \notin \hat{\mathcal{N}} \} | = 1 \} \nonumber \\
\hat{\mathcal{L}}_1 & := & \{ \ell \in [L] : | \{A_{\ell n} : n \in \hat{\mathcal{N}} \} | = 1 \}.
\end{eqnarray}
We have that $\hat{\mathcal{L}}_0 \supseteq [L] \backslash \mathcal{L}$ and $\hat{\mathcal{L}}_1 \supseteq \pi^{-1}( [L] \backslash \mathcal{L} )$, and with high probability both inclusions are actually equalities, as the following lemma shows. 

\

\textsc{Lemma.}
\begin{eqnarray}
\mathbb{P}(\hat{\mathcal{L}}_0 = [L] \backslash \mathcal{L} \ | \ \hat{\mathcal{N}} = \mathcal{N}) & = & (1-q^{1-N \bar \nu})^{L\lambda}, \nonumber \\
\mathbb{P}(\hat{\mathcal{L}}_1 = \pi^{-1}([L] \backslash \mathcal{L}) \ | \ \hat{\mathcal{N}} = \mathcal{N}) & = & (1-q^{1-N \nu})^{L\lambda}.
\end{eqnarray}

\

\textsc{Proof.}
If $\hat{\mathcal{N}} = \mathcal{N}$, we have that
\begin{eqnarray}
\hat{\mathcal{L}}_0 & = & \{ \ell \in [L] : | \{A_{\ell n} : n \notin \mathcal{N} \} | = 1 \} \nonumber \\
& = & \{ \ell \in [L] : | \{\xi^{(\mathcal{L},n)}_\ell : n \notin \mathcal{N} \} | = 1 \} \nonumber \\
& \supseteq & \{ \ell \in [L] : \forall n \notin \mathcal{N}, \xi^{(\mathcal{L},n)}_\ell = 0 \} \nonumber \\
& \supseteq & [L] \backslash \mathcal{L}. \nonumber
\end{eqnarray}
Meanwhile, $\mathbb{P} ( | \{\xi^{(\mathcal{L},n)}_\ell : n \notin \mathcal{N} \} | = 1 \ | \ \ell \in \mathcal{L} ) = q^{1-N \bar \nu}$. Similarly,
\begin{eqnarray}
\hat{\mathcal{L}}_1 & = & \{ \ell \in [L] : | \{A_{\ell n} : n \in \mathcal{N} \} | = 1 \} \nonumber \\
& = & \{ \ell \in [L] : | \{\xi^{(\mathcal{L},n)}_{\pi(\ell)} : n \in \mathcal{N} \} | = 1 \} \nonumber \\
& \supseteq & \{ \ell \in [L] : \forall n \in \mathcal{N}, \xi^{(\mathcal{L},n)}_{\pi(\ell)} = 0 \} \nonumber \\
& \supseteq & \pi^{-1}( [L] \backslash \mathcal{L} ). \nonumber
\end{eqnarray}
Finally, $\mathbb{P} ( | \{\xi^{(\mathcal{L},n)}_{\pi(\ell)} : n \in \mathcal{N} \} | = 1 \ | \ \pi(\ell) \in \mathcal{L} ) = q^{1-N\nu}$. The lemma follows by treating each $\ell \in \mathcal{L}$ as a Bernoulli trial. $\Box$

\

It will be convenient to consider partial functions, tuples, etc., where we write $\uparrow$ for an undefined value and perform formal manipulations using $\uparrow \notin \Omega$ for any set $\Omega$ and $\uparrow \ne \uparrow$.
\footnote{
Typically the formal requirements on $\uparrow$ will be satisfied by $\texttt{NaN}$ \emph{in silico}.
}
Now for $\ell \in \hat{\mathcal{L}}_0$, let $A_{\ell \hat{\mathcal{N}}}(0)$ denote the unique element of $\{A_{\ell n} : n \notin \hat{\mathcal{N}} \}$; similarly, for $\ell \in \hat{\mathcal{L}}_1$, let $A_{\ell \hat{\mathcal{N}}}(1)$ denote the unique element of $\{A_{\ell n} : n \in \hat{\mathcal{N}} \}$. For $m \in \{0,1\}$, define
\begin{equation}
\hat a^{(m)}_\ell :=
\begin{cases}
A_{\ell \hat{\mathcal{N}}}(m), & \ell \in \hat{\mathcal{L}}_m \\
\uparrow, & \ell \notin \hat{\mathcal{L}}_m.
\end{cases}
\end{equation}
Estimating $\pi$ now amounts to finding the cyclic shift $\hat \pi = (\hat L_1+1,\dots,L,1,\dots,\hat L_1)$ maximizing $|\{\ell: \hat a^{(0)}_{\hat \pi(\ell)} = \hat a^{(1)}_\ell\}|$ (i.e., aligning the partial tuples $\hat a^{(m)}$).

An obvious modification of the rearrangement inequality now permits us to treat $|\{\ell: \hat a^{(0)}_{\hat \pi(\ell)} = \hat a^{(1)}_\ell\}|$ in the same way we would the autocorrelation of a real-valued function and thereby yields the following

\

\textsc{Theorem.}
For $q > 2$ and $L$, $\lambda$, $\nu$ and $\pi$ fixed, the probability that we can recover $\mathcal{L}$, $\mathcal{N}$, and $\pi$ tends to unity exponentially in $N$. $\Box$

%
%
%
%
%
%
%
%
\begin{figure}[htbp]
\begin{center}
\includegraphics[trim = 5mm 0mm 10mm 0mm, clip, width=.75\columnwidth,keepaspectratio]{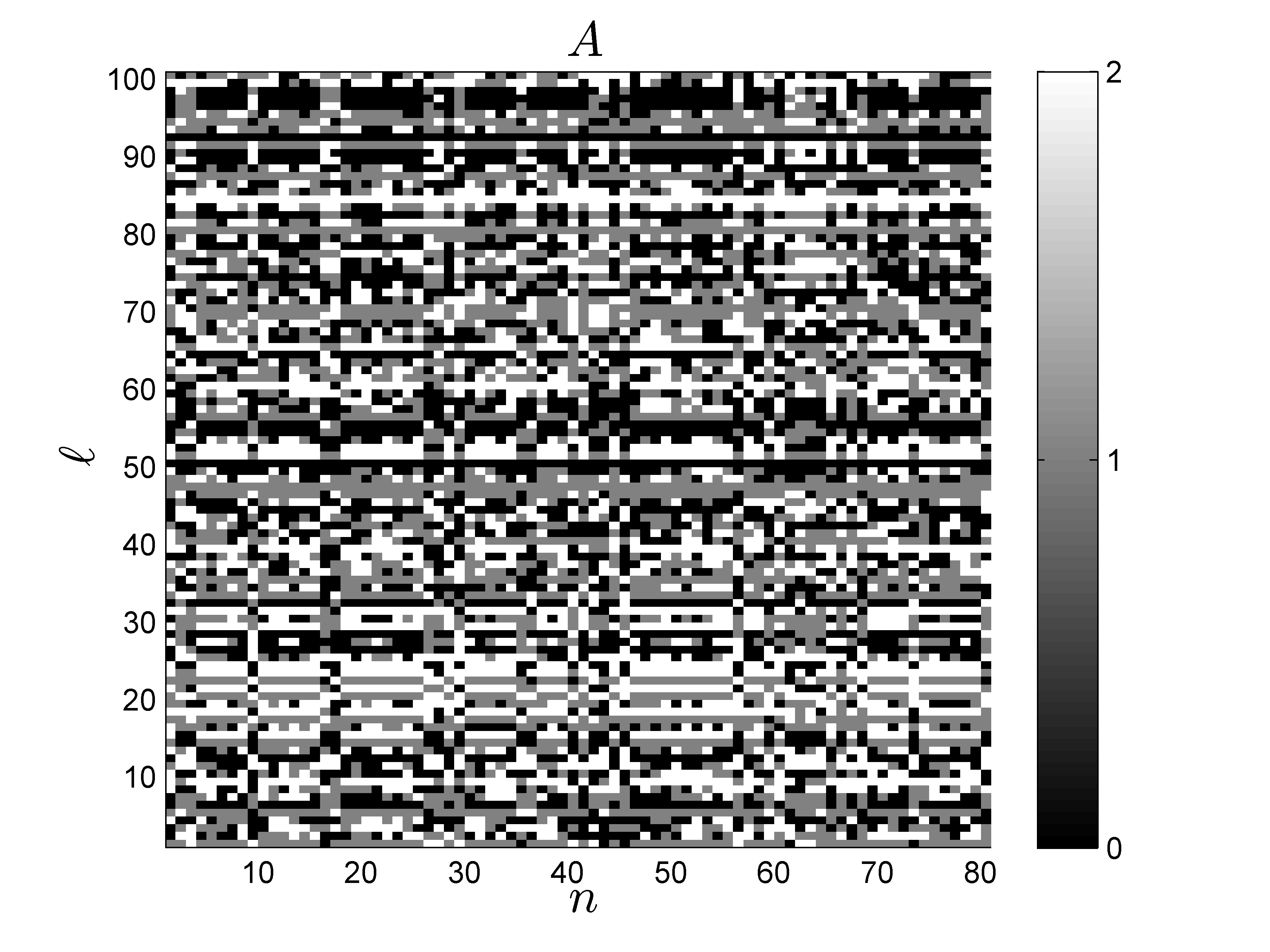}
\end{center}
\caption{ \label{fig:2unshuffle1}
2-shuffling with $q = 3$, $L = 100$, $L_1 = 40$, $N = 80$, $\lambda = \lceil 0.5 \cdot L \rceil/L$, and $\nu = \lceil 0.3 \cdot N \rceil/N$.
} 
\end{figure} %

\begin{figure}[htbp]
\begin{center}
\includegraphics[trim = 5mm 0mm 10mm 0mm, clip, width=.75\columnwidth,keepaspectratio]{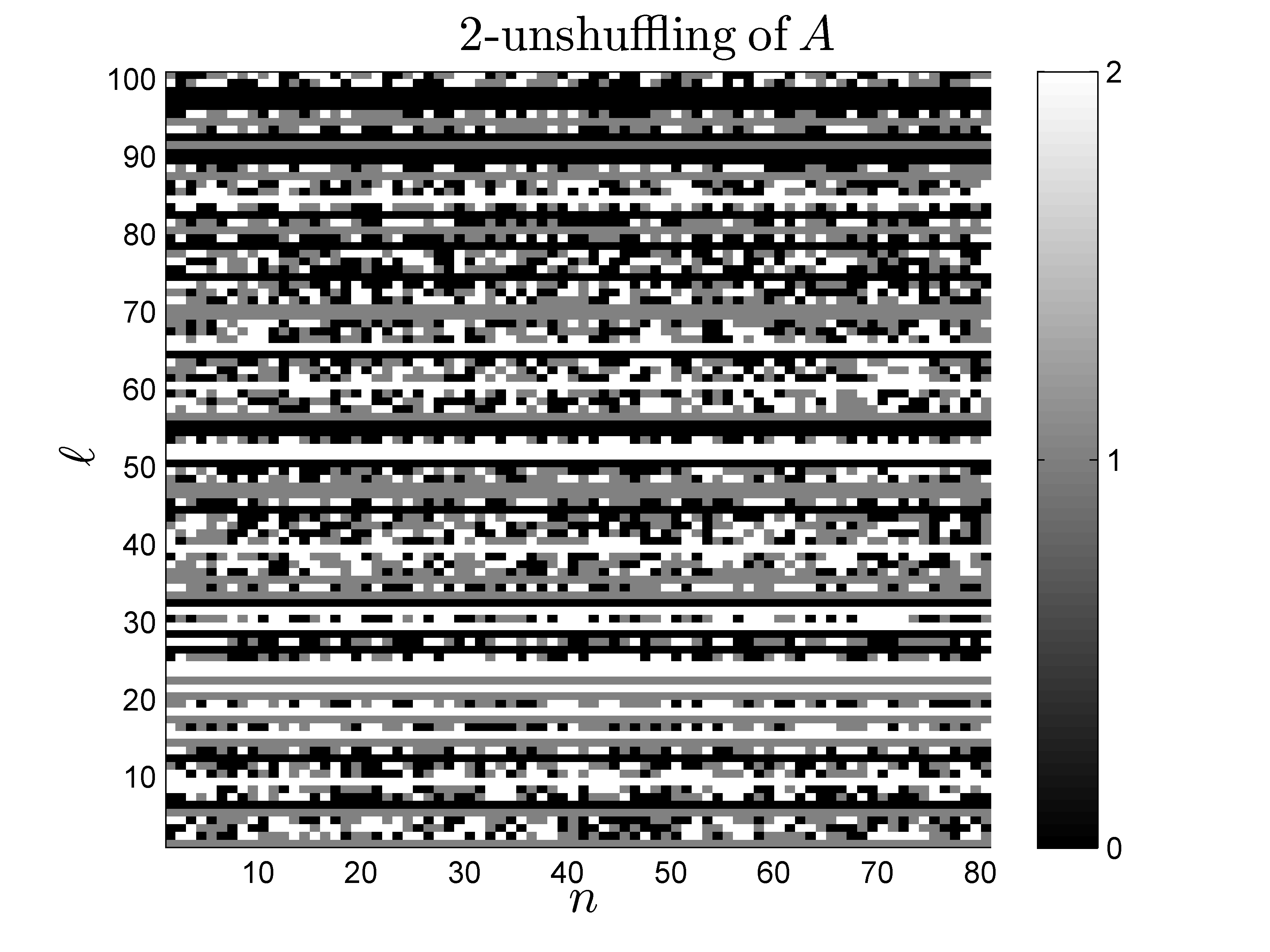}
\end{center}
\caption{ \label{fig:2unshuffle2}
Unshuffling the data in Figure \ref{fig:2unshuffle1}.
} 
\end{figure} %

\subsection{\label{sec:2UnshufflingRemarks}Remarks}

A generalization of the 2-unshuffling problem in which multiple cyclic shifts appear is possible. However, the case when all cyclic shifts are possible has already been addressed in \cite{BandeiraEtAl,Bandeira}, and residual interest therefore shifts to the question of improving the computational efficiency and/or performance of unshuffling in the case when only a small subset of all cyclic shifts are possible. Indeed, this perspective is useful, as it leads naturally into the practicalities of unshuffling a fixed class of block permutations.

\section{\label{sec:MUnshuffling}$M$-unshuffling}

\subsection{\label{sec:MUnshufflingProblem}Problem}

We use notation introduced in \S \ref{sec:Permutations}. Let $M \ge 2$ and $1 \ll L_m \in \mathbb{N}$ for $m \in [M]$ with $L := \sum_{m \in [M]} L_m$. For $\Pi : [N] \rightarrow S_M$, define $\Pi^\circledcirc_{L_1,\dots,L_M} : [N] \rightarrow S^\circledcirc_{L_1,\dots,L_M}$ by $\Pi^\circledcirc_{L_1,\dots,L_M}(n) := (\Pi(n))^\circledcirc_{L_1,\dots,L_M}$. Retaining the other conventions and definitions of \S \ref{sec:2UnshufflingProblem} (and using obvious modifications further below), define
\begin{equation}
A_{\ell n} = A^{(x,\mathcal{L},\Pi)}_{\ell n} := \left ( x+\xi^{(\mathcal{L},n)} \right )_{(\Pi^\circledcirc_{L_1,\dots,L_M}(n))(\ell)}.
\end{equation}
The \emph{general} $M$-unshuffling problem is now this: given $A$ and $M$, find $\Pi^\circledcirc_{L_1,\dots,L_M}$ (and to the extent that it matters, $\mathcal{L}$). 
\footnote{
Note that $(1,2) \circ (1_{L_1},1_{L_2}) = (1,\dots,L_1+L_2)$ and $(2,1) \circ (1_{L_1},1_{L_2}) = (L_1+1,\dots,L_1+L_2,1,\dots,L_1)$, highlighting the connection with the 2-unshuffling framework of \S \ref{sec:2UnshufflingProblem}.
}
Write $\mathcal{N}^{(\Pi)}_\sigma := \Pi^{-1}(\sigma)$ for the inverse image of $\sigma$ (not to be confused with the pointwise algebraic inverse of $\Pi$) and $\nu^{(\Pi)}_\sigma := |\mathcal{N}^{(\Pi)}_\sigma|/N$, and assume henceforth that $\Pi$ is uniformly random for $\{\nu^{(\Pi)}_\sigma\}_{\sigma \in S_M}$ fixed. We have that
\begin{equation}
A_{\ell n} = \sum_{\sigma \in S_M} 1_{\mathcal{N}^{(\Pi)}_\sigma}(n) \cdot \alpha_{\sigma^\circledcirc_{L_1,\dots,L_M}(\ell)}(n).
\end{equation}
We shall do at least one of the following to make the problem tractable:
\begin{itemize}
	\item restrict the value of one or more parameters;
	\item impose the following structural restrictions:
	\begin{enumerate}
		\item $\mathcal{L} \cap \mathcal{A} = \varnothing$, where $\mathcal{A} := \{\sum_{m' < m} L_{m'} + 1\}_{m \in [M]}$;
		\item $\mathcal{L}$ is a uniformly random subset of $[L] \backslash \mathcal{A}$ for $\lambda = |\mathcal{L}|/L$ fixed;
		\item $|\{x_\ell\}_{\ell \in \mathcal{A}}| = M$, i.e., the values $x_\ell$ are distinct for each $\ell \in \mathcal{A}$.
	\end{enumerate}
\end{itemize}
If restrictions 1 and 2 above hold, we shall speak of the \emph{restricted prefix} $M$-unshuffling problem; if in addition restriction 3 holds, we shall speak of the \emph{distinguished prefix} $M$-unshuffling problem.

It is instructive to consider a special example with $N = M!$, $\mathcal{L} = \varnothing$, and $\nu^{(\Pi)}_\sigma = N^{-1}$ identically (so that each of the elements of $S_M$ makes exactly one appearance). In this case (and by an obvious extension, more generally) the sizes of the partitions induced by the functions $n \mapsto A_{\ell n}$ behave in a way that may be surprising. For the sake of simplicity, assume that the $L_m$ are distinct and that $\{L_m\}_{m \in [M]} \subset \mathbb{N}_+$ has \emph{distinct subset sums}: i.e., $|\{\sum_{m \in S} L_m : S \subseteq [M]\}| = 2^M$ \cite{Bohman}. 
\footnote{
An open conjecture of Erd\"os is that the largest element of such a set must be at least $c2^M$ for some constant $c$.
}
Write $\{\sum_{m \in S} L_m : S \subseteq [M]\} =: \{\Lambda_j\}_{j \in [2^M]}$ with $\Lambda_{j+1} > \Lambda_j$ (so that $\Lambda_1 = 0$, $\Lambda_2 = \min_m L_m$, \dots, $\Lambda_{2^M} = \sum_{m \in [M]} L_m = L$). Now if $P(\ell)$ denotes the size of the partition induced by the function $n \mapsto A_{\ell n}$ and we define $\Lambda_0 := 0$ and $P(0) := 0$, it must be that $P(\Lambda_{j+1}) \le M + P(\Lambda_j)$, and at the same time it must be the case that $P(\ell) = P(L-\ell)$. That is, $P$ attains a maximum value of less than $M2^{M-1}$, which is much less than $M!$ for $M > 5$.

\begin{figure}[htbp]
\begin{center}
\includegraphics[trim = 5mm 0mm 10mm 0mm, clip, width=.75\columnwidth,keepaspectratio]{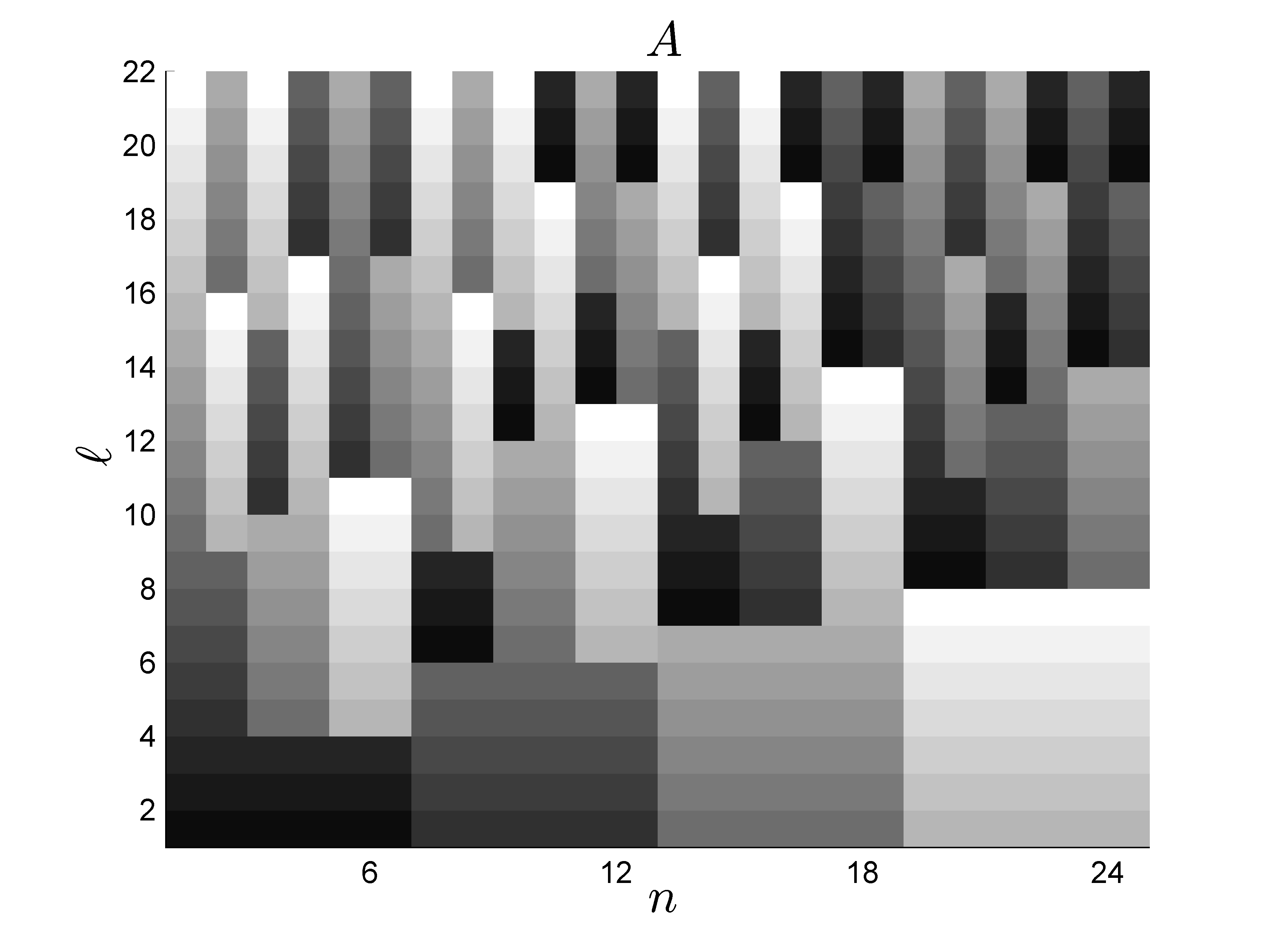}
\end{center}
\caption{ \label{fig:CleanShuffle1}
The case $M = 4$, $N = M!$, $\mathcal{L} = \varnothing$, $\nu_\sigma^{(\Pi)} \equiv N^{-1}$ and $(L_1,L_2,L_3,L_4) = (3,5,6,7)$ yields a maximum partition size of $12 \ll M!$.
} 
\end{figure} %

\begin{figure}[htbp]
\begin{center}
\includegraphics[trim = 5mm 0mm 10mm 0mm, clip, width=.75\columnwidth,keepaspectratio]{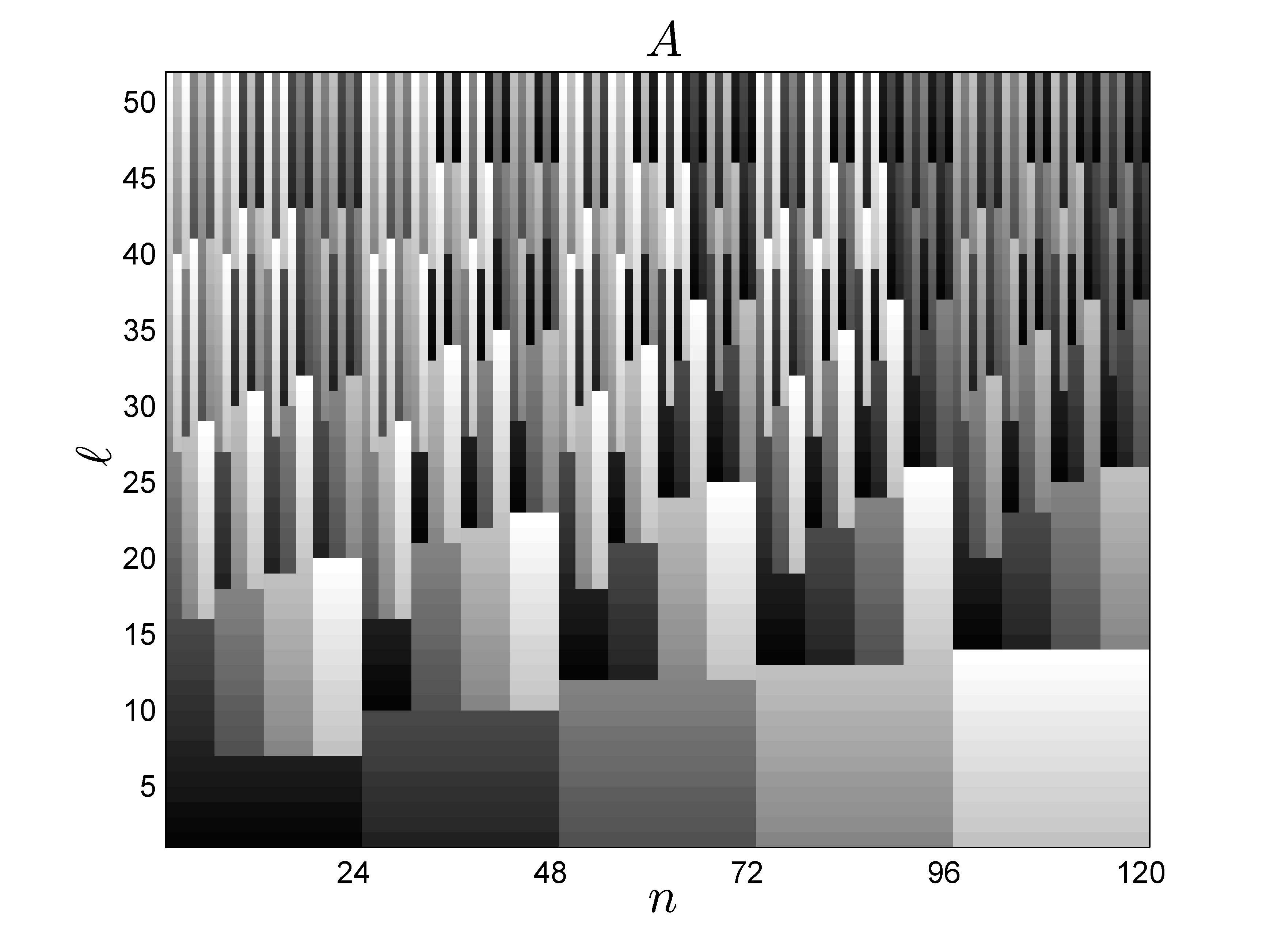}
\end{center}
\caption{ \label{fig:CleanShuffle2}
A case similar to Figure \ref{fig:CleanShuffle1} with $M = 5$ and $(L_1,L_2,L_3,L_4,L_5) = (6,9,11,12,13)$ yields a maximum partition size of $30 \ll M!$.
} 
\end{figure} %


This example suggests that we ought to focus on the local structure of shuffles, since information about their global structure is inexorably diffused among the loci $\ell$. Once this perspective is adopted, it is natural to focus on the first $\min_m L_m$ rows of $A$ in general as well as for the restricted or distinguished prefix problems, since for these first few rows there are no offsets caused by shuffling. Of course, since we do not know $\min_m L_m$, it makes further sense to focus on the smallest and/or most common partitions induced by the rows of $A$, since with high probability these will occur at or near the beginning anyway. 
We have written MATLAB code that attempts to decompose and determine the structure of shuffles according to this perspective; a general $M$-unshuffler could build upon this code to estimate $M$, $(L_m)_{m \in [M]}$, and finally motifs corresponding to blocks \emph{en route} to unshuffling via reassembly of the motifs. 

\subsection{\label{sec:SolvingRestrictedPrefixMUnshuffling}Solving the restricted prefix $M$-unshuffling problem}

Since
\begin{equation}
\sigma^\circledcirc_{L_1,\dots,L_M}(1) = \sum_{m < \sigma(1)} L_m+1
\end{equation}
in the restricted prefix case we have that
\begin{align}
A_{1n} & = & \sum_{\sigma \in S_M} 1_{\mathcal{N}^{(\Pi)}_\sigma}(n) \cdot x_{\sigma^\circledcirc_{L_1,\dots,L_M}(1)} \nonumber \\
& = & \sum_{\sigma \in S_M} 1_{\mathcal{N}^{(\Pi)}_\sigma}(n) \cdot x_{\sum_{m < \sigma(1)} L_m+1}.
\end{align}
That is, the first row of $A$ depends only on $x$ and $\Pi$. For the moment, write $f(n) := (\Pi^\circledcirc_{L_1,\dots,L_M}(n))(1)$. The usual analysis of the birthday problem shows the following

\

\textsc{Proposition.} For the restricted prefix case, the partition of $[N]$ induced by the function $n \mapsto A_{1 n}$ is refined by the partition induced by $f$, and the probability that these two partitions are identical is $p_\mathcal{N} = \binom{q}{|f([N])|} \cdot \frac{|f([N])|!}{q^{|f([N])|}} \approx e^{-|f([N])|^2/2q}$. (Note that $|f([N])| \le M$.) For the distinguished prefix case, these partitions are identical. $\Box$

\

Thus if $2q \gg |f([N])|^2$ 
\footnote{
In problems of practical interest, whenever the restricted prefix assumption holds, we expect to be able to assume this by considering successive $k$-tuples of components over $\mathbb{Z}/q\mathbb{Z}$ as individual components over $\mathbb{Z}/q^k\mathbb{Z}$.
}, 
with high probability recovering the partition of $[N]$ induced by $f$ will be easy; in the distinguished prefix case, it will always be easy. From the easy approximate bound $p_\mathcal{N}^M \gtrapprox e^{-M^3/2q}$ we get the following loosely stated

\

\textsc{Theorem.} 
Restricted prefix shuffles can be perfectly unshuffled with high probability whenever $M^3 \ll 2q$, and distinguished prefix shuffles can be perfectly unshuffled with high probability for any $M$. $\Box$

\

\begin{figure}[htbp]
\begin{center}
\includegraphics[trim = 70mm 70mm 70mm 70mm, clip, width=80mm,keepaspectratio]{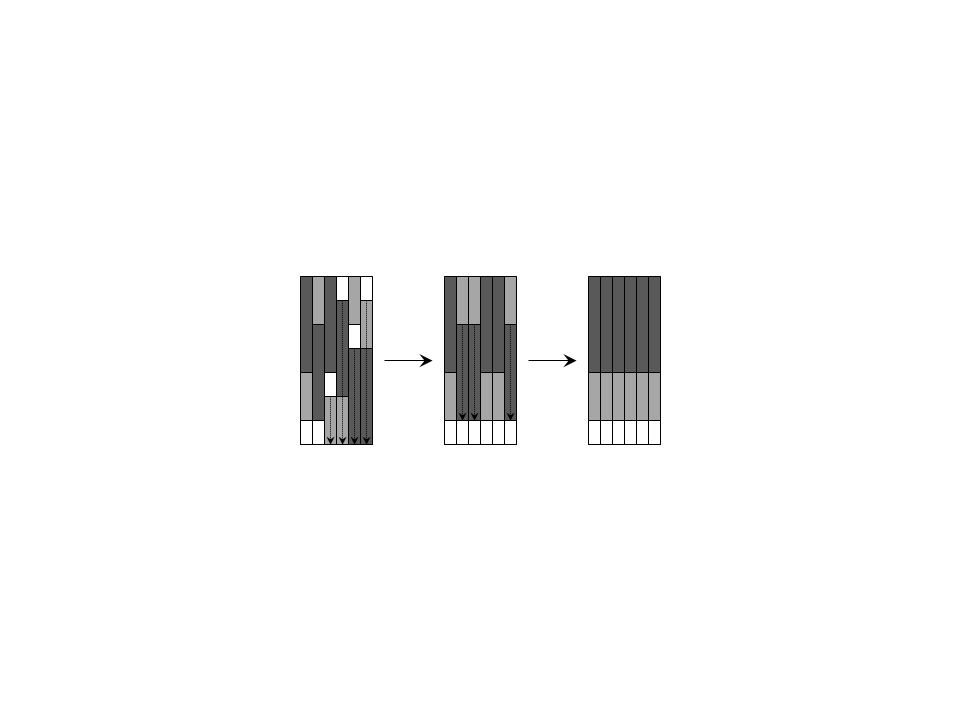}
\end{center}
\caption{ \label{fig:unshufflefig}
Schematic of a practical unshuffling technique for the restricted prefix case. Columns are aligned initially; then truncated, and the process repeats.
} 
\end{figure}

\begin{figure}[htbp]
\begin{center}
\includegraphics[trim = 5mm 0mm 10mm 0mm, clip, width=.75\columnwidth,keepaspectratio]{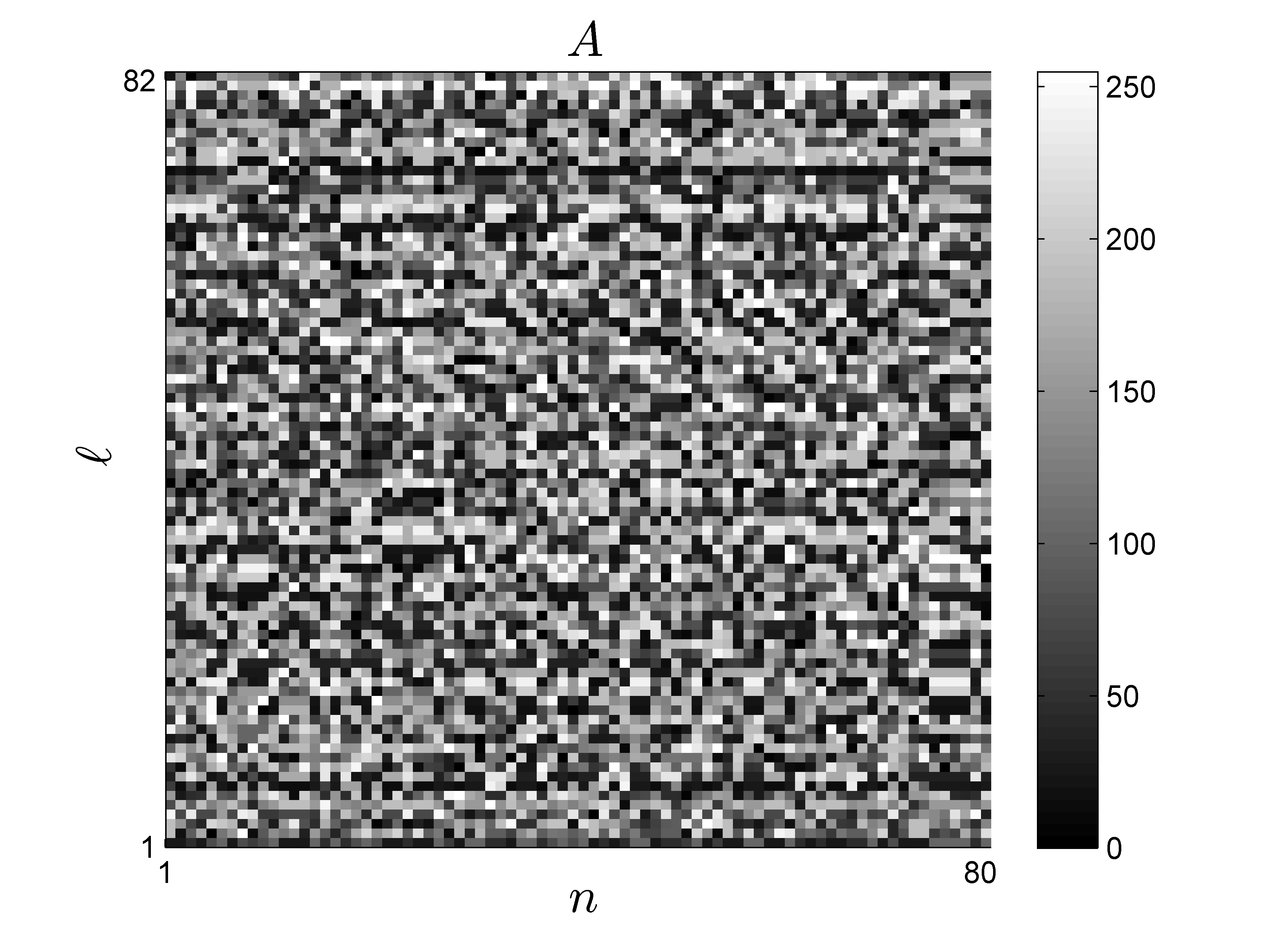}
\end{center}
\caption{ \label{fig:Munshuffle1}
6-shuffling with $q = 256$; $L_1 = 11$, $L_2 = 11$, $L_3 = 12$, $L_4 = 12$, $L_5 = 16$, $L_6 = 20$; $N = 80$; and $\lambda = \lceil 0.5 \cdot L \rceil/L$. The shuffle acts with 1 coherent block permutation (CBP; see \S \ref{sec:Permutations}) on 16 columns; 2 CBPs on 8 columns each; 4 CBPs on 4 columns each; 8 CBPs on 2 columns each; and 16 CBPs on individual columns. That is, 31 of the 720 possible CBPs are realized.
} 
\end{figure}

\begin{figure}[htbp]
\begin{center}
\includegraphics[trim = 5mm 0mm 10mm 0mm, clip, width=.75\columnwidth,keepaspectratio]{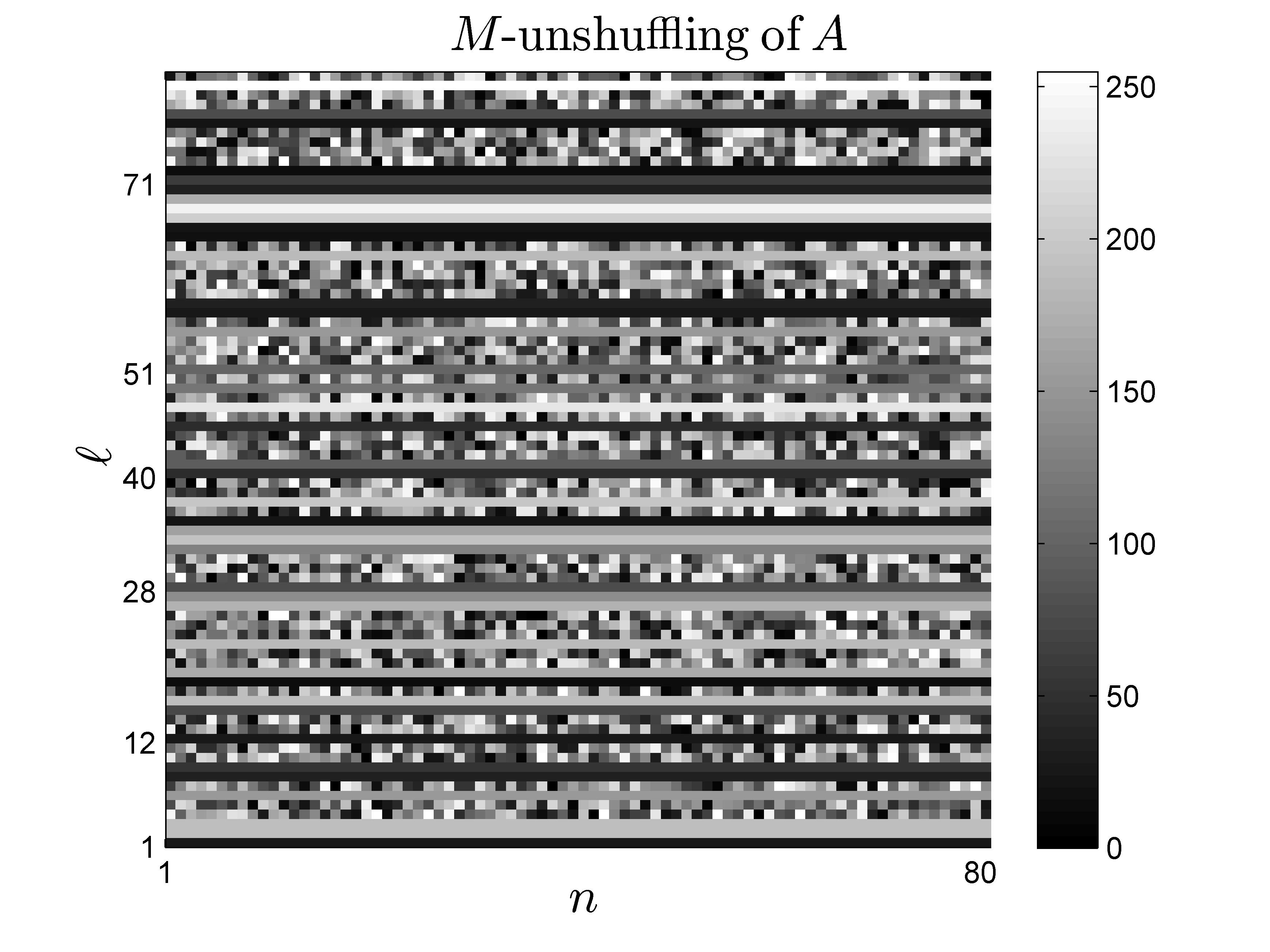}
\end{center}
\caption{ \label{fig:Munshuffle2}
Unshuffling the data in Figure \ref{fig:Munshuffle1} succeeds, and also recovers $M = 6$ and the block lengths $(L_m)_{m \in [M]}$. Note that $M^3 = 216 \ll 512 = 2q$. Decreasing $q$ to (say) 128 but otherwise performing the same analysis (including the same pseudorandom seed) still turns out to yield perfect reconstruction (\emph{a priori} this is to be expected with probability at least 0.43), but $q = 64$ fails spectacularly, reconstructing only the first two blocks correctly before yielding the estimate $\hat M = 24 \gg 6$.
} 
\end{figure}

To achieve unshuffling in practice for the restricted prefix case, a simple modification of the technique for finding $\pi$ described at the end of \S \ref{sec:2UnshufflingPi} 
suffices. The idea (see figure \ref{fig:unshufflefig}) is to circularly shift each column of $A$ so that the first rows align and to iterate this process on the matrix obtained by eliminating the aligned rows. By the restricted prefix assumption, it is not necessary to perform the sort of analysis of partitions used for unshuffling an unrestricted 2-shuffle: instead, in our implementation we merely incorporate a weight of $2^{-\ell}$ for aligning the $\ell$th row, i.e., the $k$th column is aligned to (say) the first via the circular shift
\begin{equation}
\arg \max_{\hat \pi \in \mathbb{Z}/L\mathbb{Z} \hookrightarrow S_L} \sum_{\ell = 1}^L 2^{-\ell} \delta_{A_{\ell 1},A_{\hat \pi(\ell) k}},
\end{equation}
where as usual $\delta$ denotes the Kronecker delta. The weight of $2^{-\ell}$ ensures that the first row contributes slightly more to the alignment than all the remaining rows combined, dovetailing with the restricted prefix assumption. Figures \ref{fig:Munshuffle1} and \ref{fig:Munshuffle2} show this technique in action.

%
%
%
%
%
%
%
%
%
%

\section{\label{sec:SynchronizationProblems}Synchronization problems}

Unshuffling can be viewed as a so-called \emph{synchronization problem} \cite{Bandeira}. Consider a finite graph $G = (V,E)$, a compact group $\mathcal{G}$, and let $f: E \times \mathcal{G} \rightarrow \mathbb{R}$. The synchronization problem corresponding to $G$, $\mathcal{G}$, and $f$ is to find a \emph{potential} $g: V \rightarrow \mathcal{G}$ minimizing 
\begin{equation}
\label{eq:SynchObj}
\sum_{(j,k) \in E} f((j,k), g_j g_k^{-1}).
\end{equation}
An interesting variant of the synchronization problem arises upon requiring $g(V) \subseteq \mathcal{H} \subset \mathcal{G}$ for $| \mathcal{H} | \ll | V |$. 

As an example, let
\begin{itemize}
	\item $G = K_N$ (i.e., the complete graph on $N$ vertices);
	\item $\mathcal{H} \subset \mathcal{G} = S_L$ be the set of coherent block permutations (see \S \ref{sec:Permutations}) of the form $\sigma^\circledcirc_{L_1,\dots,L_M}$ for some fixed $\{L_m\}_{m=1}^M$ that is completely unknown apart from the implicit constraint $L = \sum_{m=1}^M L_m$;
	\item $f((j,k), (\sigma_j)^\circledcirc_{L_1,\dots,L_M} ((\sigma_k)^\circledcirc_{L_1,\dots,L_M})^{-1}) = -\langle y_j, \rho((\sigma_j)^\circledcirc_{L_1,\dots,L_M}((\sigma_k)^\circledcirc_{L_1,\dots,L_M})^{-1})^* y_k \rangle$, where $\rho : S_L \rightarrow GL(L)$ is the natural permutation representation: $\rho(\sigma)_{ab} := \delta_{b,\sigma(a)}$.
\end{itemize}

At this point it is probably best to unpack the preceding paragraph by furnishing a somewhat more concrete point of view that leads to it. Suppose that we have a fixed set $\{x^{(m)}\}_{m=1}^M$ of tuples with $\dim x^{(m)} = L_m$. Our only piece of information about this set is the value of $L = \sum_{m=1}^M L_m$. We are given $N$ samples of the form $y_j = R_j (x^\oplus + \xi_j^\oplus)$, where $R_j := \rho((\sigma_j)^\circledcirc_{L_1,\dots,L_M})^*$, $x^\oplus := \oplus_m x^{(m)}$ and the implied $\xi_j^{(m)}$ are random variables that are IID w/r/t $j$. That is, the $x^{(m)}$ have noise added and are then shuffled. To unshuffle the $y_j$, we must find the coherent block permutations that minimize
\begin{equation}
\label{eq:SynchObj2}
-\sum_{j,k} \langle R_j^* y_j, R_k^* y_k \rangle.
\end{equation}
But since $z^*Az = \text{Tr}(Azz^*)$ and 
\begin{equation}
\sum_j R_j^* y_j = \begin{pmatrix}R_1^* & \dots & R_N^*\end{pmatrix} \begin{pmatrix} y_1 \\ \vdots \\ y_N \end{pmatrix}
\end{equation}
the objective function in \eqref{eq:SynchObj2} equals
\begin{equation}
\label{eq:SynchObj3}
- \text{Tr} \left( \begin{pmatrix} R_1 \\ \vdots \\ R_N \end{pmatrix} \begin{pmatrix}R_1^* & \dots & R_N^*\end{pmatrix} \begin{pmatrix} y_1 \\ \vdots \\ y_N \end{pmatrix} \begin{pmatrix}y_1^* & \dots & y_N^*\end{pmatrix} \right )
\end{equation}
which can be written as $\text{Tr}(RY)$ using an obvious shorthand notation, reflecting the typical formulation of synchronization as a semidefinite program.

This problem presents interrelated difficulties beyond those encountered in the synchronization problems studied in \cite{Bandeira} and \cite{BandeiraEtAl}: first, $\mathcal{G} = S_L$ is large but discrete; and second, the subset of values that a potential can take is complicated. However, one slightly mitigating observation is that we may try multiple values of $M$ and determine the correct one \emph{a posteriori}, so we may assume $M$ is known.

\appendices

\section{\label{sec:Permutations}Permutations}

Let $S_M$ denote the symmetric group on $M$ elements and write $\sigma = (\sigma(1),\dots,\sigma(M)) \in S_M$. Now for $L := \sum_{m=1}^M L_m$ and $\tau^{(m)} \in S_{L_m}$ for $m \in [M]$, define the \emph{(permutation) operad composition} 
\footnote{
See, e.g. \S 2.2.20 of \cite{Leinster}.
}
$\circ : S_M \times \prod_{m=1}^M S_{L_m} \rightarrow S_L$ as follows:
\begin{equation}
\label{eq:PermOperad1}
\circ : (\sigma, \tau^{(1)}, \dots, \tau^{(M)}) \mapsto \sigma \circ (\tau^{(1)}, \dots, \tau^{(M)}),
\end{equation}
where $\sigma \circ (\tau^{(1)}, \dots, \tau^{(M)})({\textstyle \sum_{m<n} L_m} + \ell) := \sum_{m<\sigma(n)} L_{\sigma^{-1}(m)} + \tau^{(n)}(\ell)$.
For notational convenience, write $1_M := (1,\dots,M)$ for the identity of $S_M$. The \emph{block permutation} $\sigma_{L_1,\dots,L_M} \in S_L$ is 
\footnote{
See, e.g. \S 1.2 of \cite{MarklShniderStasheff} or definition 8.1.4 of \cite{Kitaev}.
}
\begin{equation}
\label{eq:BlockPerm}
\sigma_{L_1,\dots,L_M} := \sigma \circ (1_{L_1}, \dots, 1_{L_M}).
\end{equation}
For example, $(4,2,1,3)_{4,3,3,2} = (4,2,1,3) \circ (1_4,1_3,1_3,1_2) = (9,10,11,12,4,5,6,1,2,3,7,8)$. We shall write $S_{L_1,\dots,L_M} \subseteq S_L$ for the set of block permutations of the form \eqref{eq:BlockPerm}. It is easy to show that
\begin{equation}
(\sigma_{L_1,\dots,L_M})^{-1} = (\sigma^{-1})_{L_{\sigma^{-1}(1)},\dots,L_{\sigma^{-1}(M)}}.
\end{equation}

Define the \emph{coherent block permutation} 
\begin{equation}
\label{eq:CoherentBlockPerm}
\sigma^\circledcirc_{L_1,\dots,L_M} := \sigma \circ (1_{L_{\sigma(1)}}, \dots, 1_{L_{\sigma(M)}}) = \sigma_{L_{\sigma(1)},\dots,L_{\sigma(M)}}.
\end{equation}
For example, $(4,2,1,3)^\circledcirc_{4,3,3,2} = (4,2,1,3) \circ (1_2,1_3,1_4,1_3) = (11,12,5,6,7,1,2,3,4,8,9,10)$. We shall write $S^\circledcirc_{L_1,\dots,L_M} \subseteq S_L$ for the set of block permutations of the form \eqref{eq:CoherentBlockPerm}. The crucial property of $S^\circledcirc_{L_1,\dots,L_M}$ is that its elements permute the intervals $\{\sum_{m'<m} L_{m'} + 1,\dots,\sum_{m' \le m} L_{m'}\}$ of $[L]$ (note that coherent block permutations are therefore ``integral'' \emph{interval exchange transformations} \cite{AvilaForni}), or more formally that the functions $m \mapsto \sum_{m'<m} L_{m'} + 1$ and $m \mapsto \sum_{m' \le m} L_{m'}$ are equivariant with respect to the natural action of $S_M$ on $[M]$ and the corresponding action by coherent block permutations on $[L]$. $S^\circledcirc_{L_1,\dots,L_M}$ has more complicated algebraic structure than $S_{L_1,\dots,L_M}$: to wit, we have that
\begin{equation}
\sigma^\circledcirc_{L_1,\dots,L_M} = ((\sigma^{-1})_{L_1,\dots,L_M})^{-1}.
\end{equation}

\section*{Acknowledgment}

We thank Afonso Bandeira for useful conversations.

\section*{Competing Interests}

Declarations of interest: none.

\end{document}